\documentclass{jfm}
\usepackage{graphicx}
\usepackage{subfigure}
\usepackage{color}

\begin{document}

\newtheorem{lemma}{Lemma}
\newtheorem{corollary}{Corollary}

\shorttitle{Inertial flow around obstacles in microchannels} 
\shortauthor{Hamed Haddadi} 

\title{Inertial flow around obstacles in microchannels}

\author
 {
Hamed Haddadi \thanks{Email address for correspondence: haddadi@ucla.edu}\ns \break
  }

\affiliation
{
\aff{1}
Department of Bioengineering, University of California Los Angeles
}

\maketitle

\begin{abstract}
Formation of recirculating wakes is a prominent feature of inertial flow around bluff bodies. Below the onset of vortex shedding in uniform unbounded flows, the fluid in the recirculating wake region moves on closed planar orbits. The steady wake is thus an isolated zone in the flow and does not exchange fluid with the free stream. In this work, we utilize lattice-Boltzmann simulations and microfluidic experiments to demonstrate that in microchannel inertial flow of Newtonian fluids, the recirculating wake is replaced by a three-dimensional vortical flow. Spiraling streamlines generate a continuous exchange of fluid between the vortex behind the obstacle and the free stream. The flow inertia is represented by Reynolds number defined as $Re = \frac{u_{max}D_{y}}{\nu}$, where $u_{max}$ is the maximum fluid velocity in the channel inlet, $D_{y}$ is the characteristic obstacle length and $\nu$ is the fluid kinematic viscosity. We discuss the effects of $Re$, the obstacle shape and the wall confinement on the fluid entry into the vortex. Further, we demonstrate that in flow of a dilute suspension of particles around the obstacle, the fluid entry into the vortex can result in entrapment of particles as well.  
\end{abstract}

\section{Introduction}

In this work we utilize lattice-Boltzmann (LBM) simulations and microfluidic experiments to address the flow of Newtonian fluids and dilute suspensions around bluff bodies in microchannels. We focus on the exchange of fluid and particles with the vortex behind the obstacle in the pressure-driven flow in microchannels. \newline 

Formation of recirculating wakes is a well-known feature of inertial flow around bluff bodies. Reynolds number $Re$ represents the flow inertia which is defined using a characteristic length $L_{c}$, a characteristic velocity $U_{c}$ and the kinematic viscosity of the fluid $\nu$ as $Re = \frac{L_{c}U_{c}}{\nu}$.  Around $Re \approx O(10)$, streamlines close to the forward face of the body leave the body and pass around a recirculating wake. With increasing $Re$, the size of the recirculating wake increases until the instabilities interrupt the steady flow. The instabilities lead to the vortex shedding phenomenon which appears in the form asymmetric oscillation of two wakes behind the obstacle. The rotating fluid entrapped in the steady recirculating wake is periodically released and moves downstream, leading to formation of ``vortex-streets" (Batchelor 1967). Before the onset of vortex shedding, a limiting streamline, i.e the separatrix, specifies the boundary between the recirculating flow and the free stream.  \newline

Uniform flow of Newtonian fluids around the obstacle is a classic fluid dynamics problem. However, the presence of walls alters several physical aspects of the flow. For instance, the drag and lift forces on the obstacle are different than uniform flow due to parabolic velocity profile between two parallel walls. The presence of walls impedes the vortex shedding and the unsteady flow appears at higher $Re$.  As an example, the onset of vortex shedding for flow around a circular cylinder placed symmetrically between two parallel walls has been observed at $Re = 200$, higher than $Re = 55$ for uniform unbounded flows (Zovatto \& Pedrizzetti 2001). \newline

The motivation of the present study is our prior experiments of dilute suspension flow around obstacles in a microchannel (Haddadi \emph{et al.} 2014; Haddadi, Shojaei-Zadeh \& Morris 2016). Using high speed imaging of dilute suspension flow in a straight microchannel, we observed formation of a depleted wake region behind the circular and square cylinders for almost all values of $Re$. Formation of a depleted region behind the circular cylinder has also been studied for intermediate concentrations using a suspension balance model combined with the immersed boundary method (Dbouk 2016). We observed that the state of the wake depends on the obstacle shape. As opposed to the circular and square cylinders, the entry of particles into the wake was significantly higher for a narrow rectangular cylinder. In the first step, we investigated dynamics of inertial particles in uniform unbounded flows around a circular cylinder. An isolated particle released in the uniform flow does not enter the wake and passes around the obstacle. The particle released in the recirculating zone migrates outward and approaches a limit cycle trajectory close to the wake boundaries (Haddadi \emph{e al.} 2014).  By implementing a numerical model for collision of particles with sharp corners, we studied the uniform flow around square cylinders as well (Haddadi, Shjaei-Zadeh \& Morris  2016). In the latter work, we briefly mentioned formation of flow spirals from the channel into the vortex in a wall-confined microchannel flow. The flow spirals can potentially augment the entry of particles into the vortex. In the present work, we utilize LBM simulations and microfluidic experiments to study the exchange of fluid and particles with the vortex behind the obstacle in pressure-drive microfluidic flows. \newline

\section{Problem description}
We study inertial flow of Newtonian fluids and dilute suspensions of hard spherical particles around obstacles in microchannels. We initially investigate the effects of $Re$, obstacle shape and wall-confinement on entry of fluid into the vortex behind the obstacle.  Later, we discuss the exchange of particles with the vortex in flow of dilute suspensions.   \newline

We study the flow in a microchannel of length $L$, width $W$ and height $H$ along the flow, primary and secondary directions of velocity gradient respectively. The flow direction is specified by $X$ and the primary and secondary directions of velocity gradient by $Y$ and $Z$, respectively. We study the wall confinement by changing the channel height in $Z$ direction. The obstacle center of mass is symmetrically placed at $x = L/2$ and $y = W/2$ and spans the entire channel height $z = H$. We study cylindrical obstacles with circular, square and narrow rectangular cross-sectional shapes which we term the ``blade" for the rest of our discussion. Our numerical and experimental results show that the state of the wake behind the square and circular cylinders are similar. So we mainly focus on square and blade cylinders. We characterize the effect of obstacle geometry using the aspect ratio and the blockage ratio. The aspect ratio of rectangular obstacles $\lambda = \frac{D_x}{D_y}$ is representative of the obstacle shape, where $D_x$ and $D_y$ denote the lengths in flow and the primary direction of the velocity gradient respectively.  \newline

The blockage ratio is defined as $\beta = \frac{D_y}{W}$. In this work, we study three blockage ratios of the blade obstacle: $\beta = 0.25$, $0.5$ and $0.75$. Considering that we change the blockage ratio $\beta$ by increasing $D_{y}$ at constant values of $D_{x}$, the aspect ratio of the obstacle $\lambda$ changes as well. The aspect ratios for the $\beta = 0.25$, $0.5$ and $0.75$ blade cylinders correspond to $\lambda = 0.0625$, $0.0312$ and $0.0208$.  For inertial flow of a Newtonian fluid with kinematic viscosity $\nu$ inside a wall-confined conduit, $Re$ has been conventionally defined using the average inlet velocity and the hydraulic diameter of the channel  as $h_d = \frac{4A}{P}$ where $A$ and $P$ indicate the area and perimeter of the channel cross-section respectively. Here, we use the maximum inlet velocity of the fluid $u_{max}$ and the characteristic obstacle length $D_{y}$ facing the flow to define $Re$. This definition provides a better criterion for comparing the channels with different heights. Obviously, the results are invariant of $Re$ definition for channels of identical size. \newline

We use the same computational and experimental methods that were explaied in our prior works (Haddadi \emph{et. al} 2014; Haddadi, Shojaei-Zadeh \& Morris 2016; Haddadi \& Di Carlo 2017). We briefly mention our computational parameters. We conduct our simulations in a rectangular box which is periodic in the flow direction $X$ and bounded by walls in $Y$ and $Z$ directions. The sizes of the computational box in the flow and the primary velocity gradient directions are $X_{b} \times Y_{b} = 1024 \times 128$ lattice units (l.u). We studied the effect of wall confinement for two values of $Z_{b} = 64$ and $32$ l.u.  For simulations of the suspension flow, the particle diameter is $d = 5.848$ l.u.  Other computational parameters are similar to our previous works. We use $\frac{D_y}{2} = 32$ for a $\beta = 0.5$ obstacle to scale dimensions. \newline  

We organized the manuscript as follows: In section \S 3, we discuss the spiraling flow of fluid into a vortex behind the obstacle in flow of Newtonian fluids around the obstacle in microchannels. In section \S 4, we explain the exchange of particles with vortices. The experimental results will be presented in section \S 5. Finally, we present the concluding remarks in \S 6. \newline

\begin{figure}
\centering
\subfigure[$Re = 30$]{\includegraphics[totalheight=0.155\textheight,]{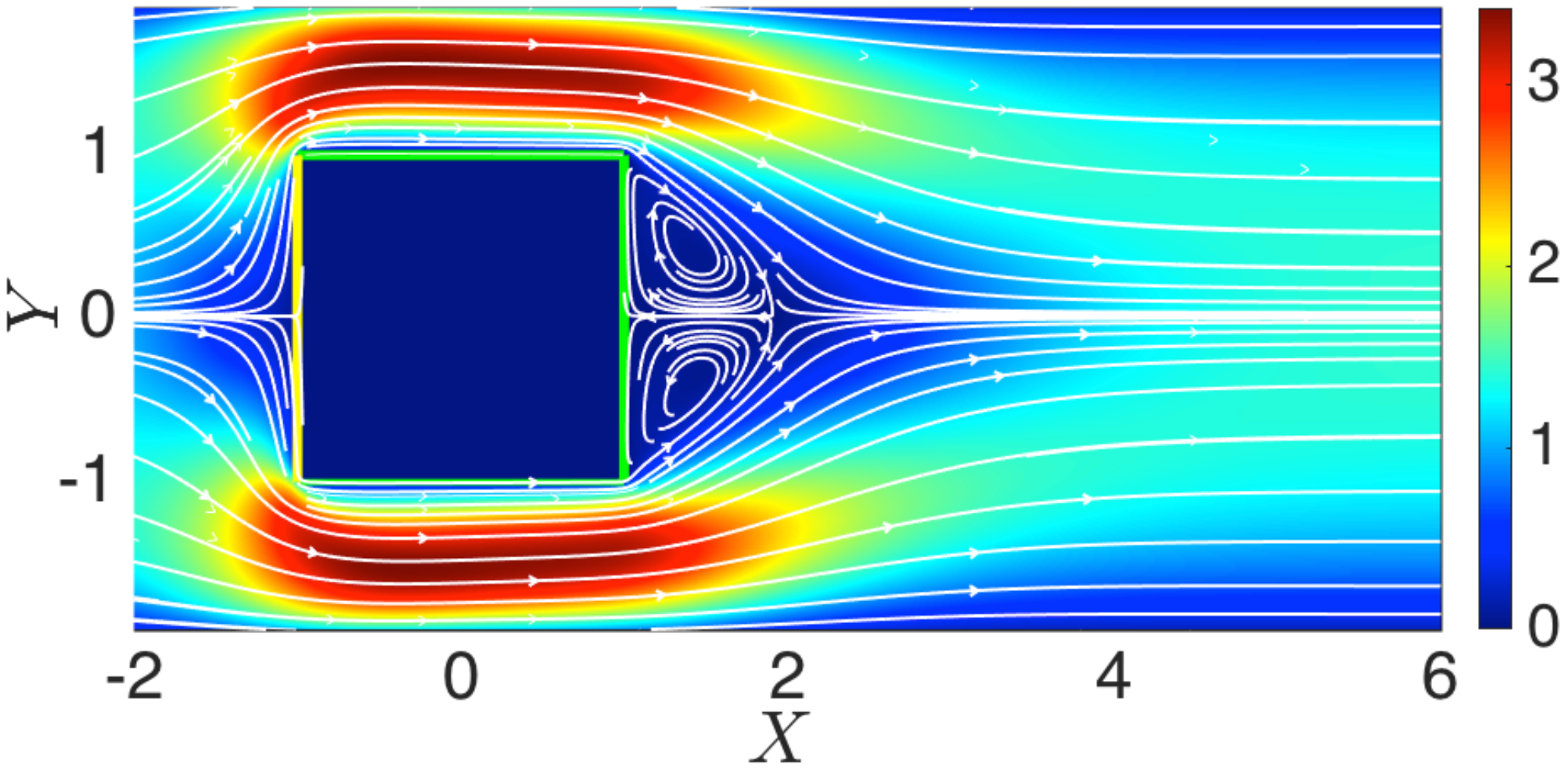}}
\subfigure[$Re = 240$]{\includegraphics[totalheight=0.155\textheight,]{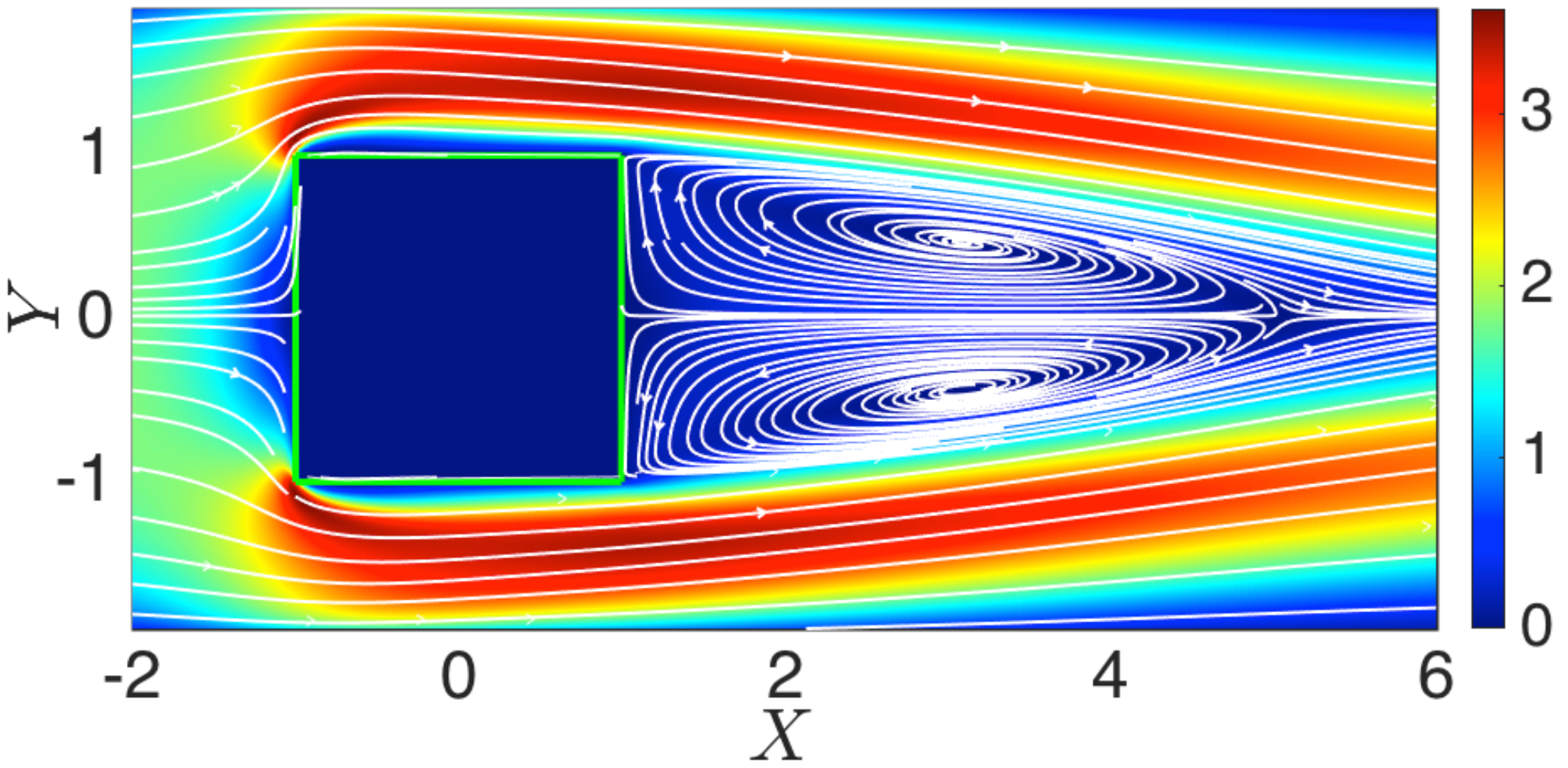}}
\caption{The longitudinal size of the wake behind a square cylinder at (\emph{a}) $Re = 30$ and (\emph{b}) $Re = 240$. The channel sizes in $X$, $Y$ and $Z$ directions are $H = 32$, $W = 4$ and $H = 2$ (using $\frac{D_{y}}{2} = 32$ l.u of a $\beta = 0.5$ obstacle as the length scale). The $\beta= 0.5$ obstacle is located in the middle of the channel and spans the entire height in $Z$ direction. The streamlines on $Z = 1$ plane are presented. The longitudinal vortex size does not change on different $Z$ planes. }
\label{fig:SDF}
\end{figure}

\begin{figure}
\centering
\subfigure[]{\includegraphics[totalheight=0.11\textheight,]{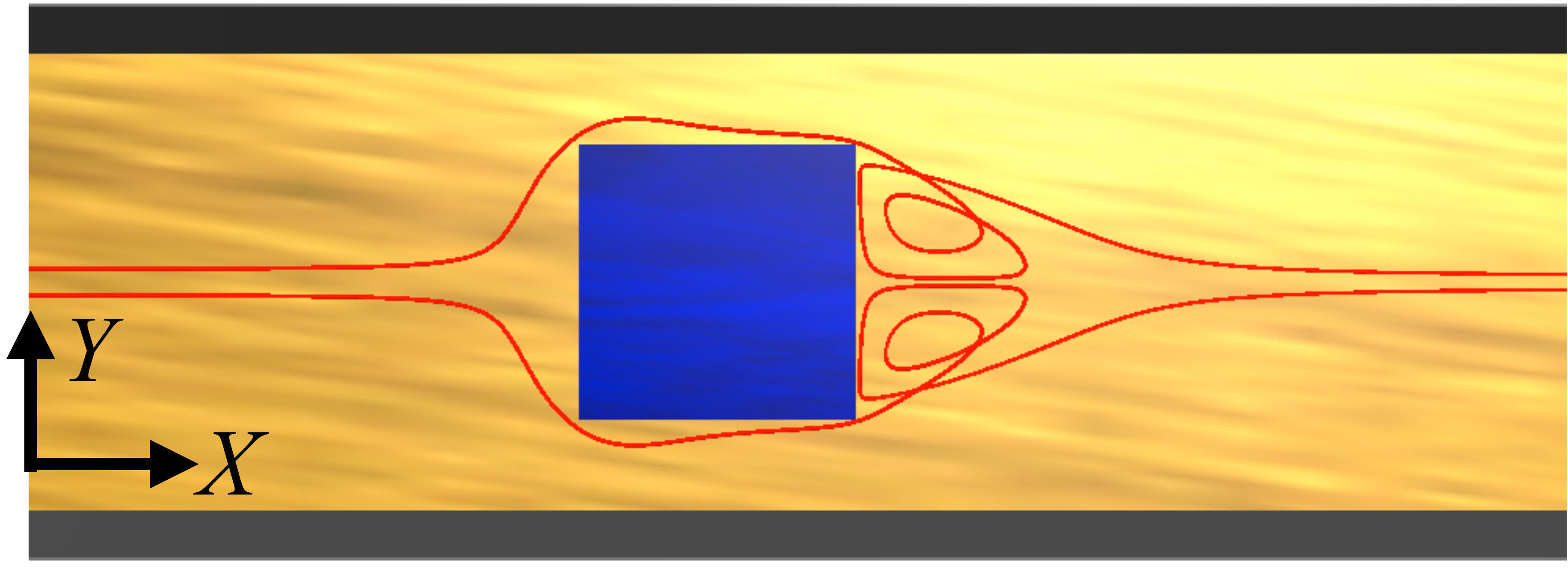}}
\subfigure[]{\includegraphics[totalheight=0.11\textheight,]{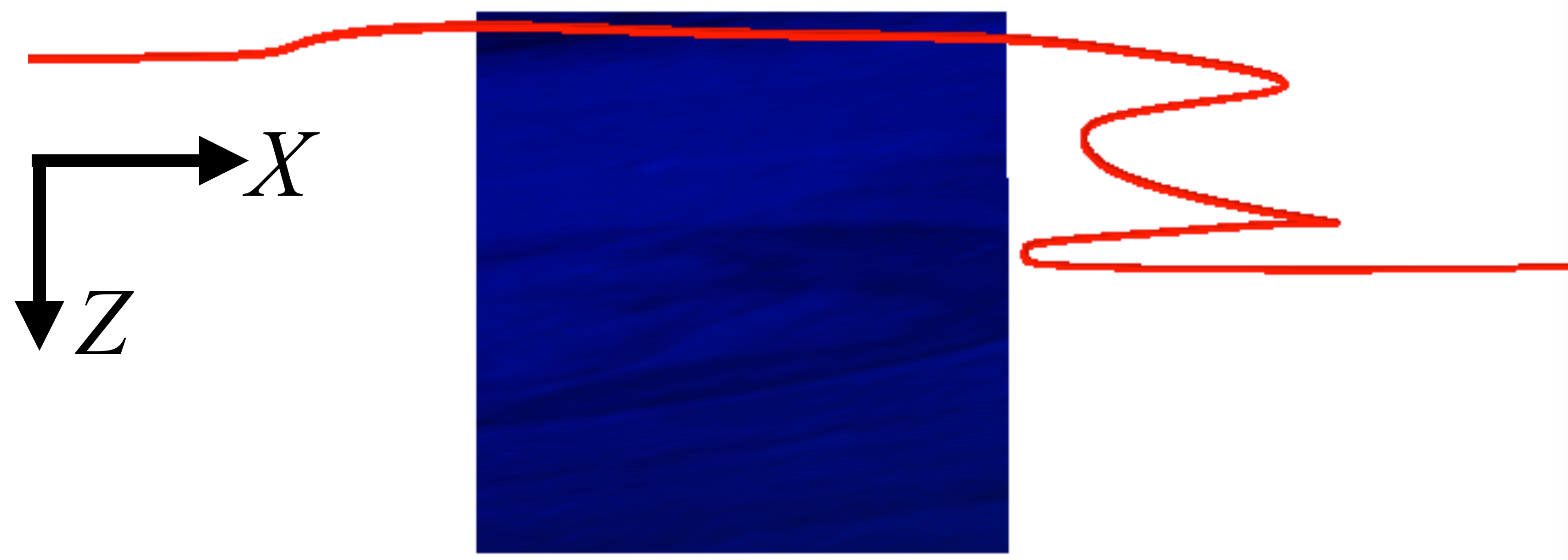}}
\caption{Two views of the sample spiraling streamlines entering in the vortex. The streamlines spiral towards the middle $Z$ plane and leave the vortical flow towards channel downstream. }
\label{fig:Entry}
\end{figure}

\section{Fluid flow around obstacles in microchannels}
Figure~\ref{fig:SDF} exhibits the longitudinal size of the vortex behind a $\beta = 0.5$ square obstacle ($\lambda = 1$) at $Re = 30$ and $240$. Similar to uniform flow around an obstacle, the longitudinal vortex size increases with $Re$. We demonstrate the streamlines on $Z = 1$ plane in the middle of the channel. We remind that lengths are normalized by $\frac{D_{y}}{2} = 32$ of a $\beta = 0.5$ obstacle. The longitudinal vortex size is identical in all $Z$ planes, including the planes adjacent to $Z = 0$ and $2$ walls. \newline  

 \begin{figure}
\centering
\subfigure[$Re = 30$]{\includegraphics[totalheight=0.12\textheight,]{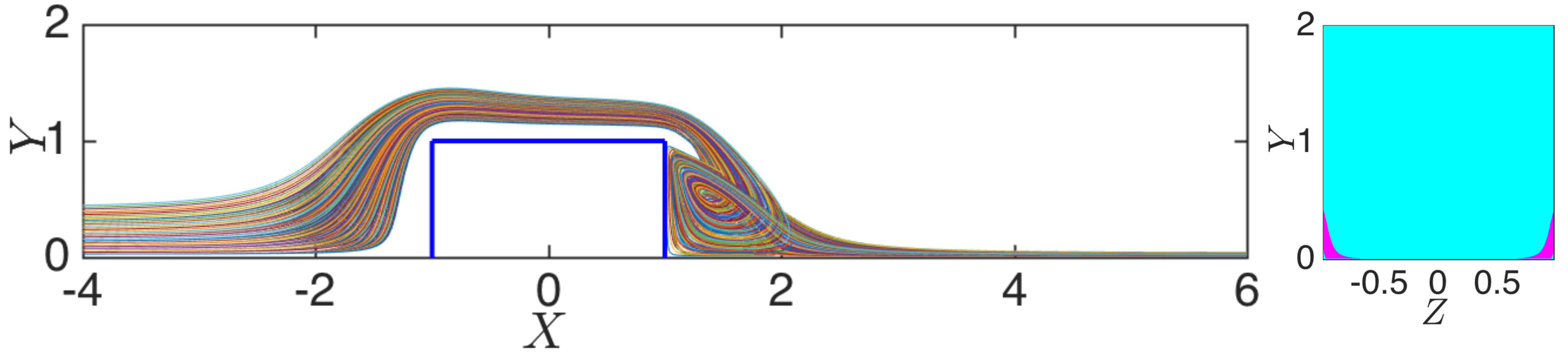}}
\subfigure[$Re = 120$]{\includegraphics[totalheight=0.12\textheight,]{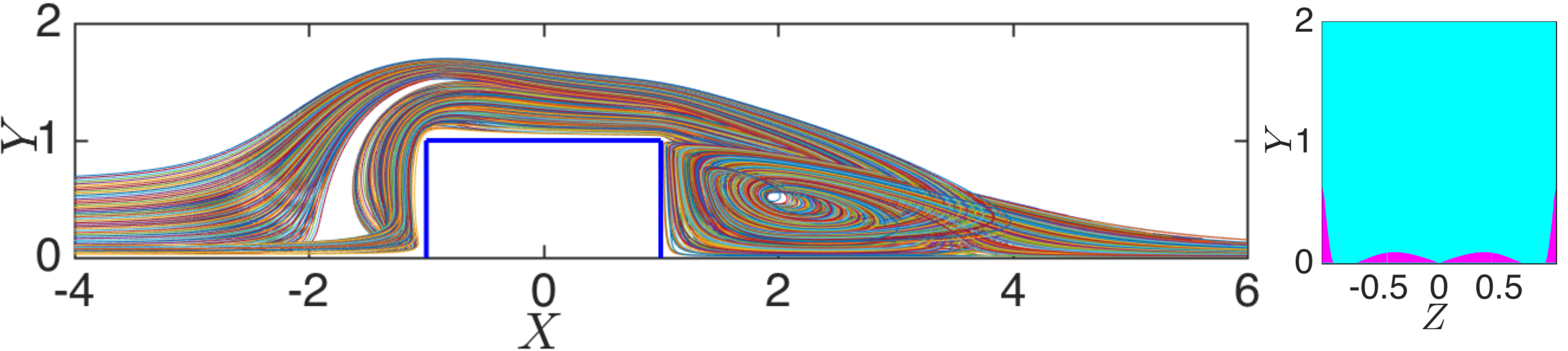}}
\subfigure[$Re = 240$]{\includegraphics[totalheight=0.12\textheight,]{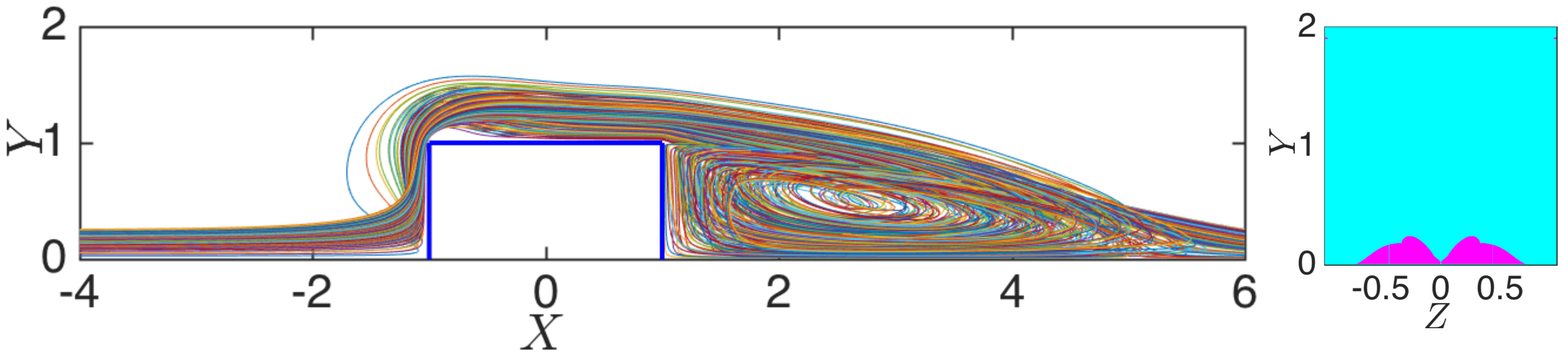}}
\caption{The basin of streamlines that spiral into a vortex behind the obstacle. Formation of the vortex is a clear indication that the notion of separatrix for unbounded flows does not apply to flow around bluff bodies in microchannels. Swirling streams of fluid transfer the mass into a vortical flow region behind the obstacle.  }
\label{fig:Basin32}
\end{figure}

To discuss the exchange of fluid with the wake behind the obstacle, we initially show sample streamlines that spiral into the vortex behind a square obstacle in figure~\ref{fig:Entry}. Two views of the same streamline clearly show the three-dimensional, spiraling flow of fluid into a vortical flow region behind a $\beta = 0.5$ square cylinder. As opposed to planar two-dimensional streamlines in the recirculating wakes in uniform flows, the streamlines in a wall-confined obstacle enter the vortex, spiral inside the vortex and leave towards downstream. The exchange of fluid mass does not occur in steady, unbounded flow of Newtonian fluids around a bluff body. Below the onset of vortex shedding, the recirculating wake is isolated from the free stream and the separatrix streamline specifies a boundary between the recirculation flow and the free stream.  \newline

\begin{figure}
\centering
\includegraphics[totalheight=0.24\textheight,]{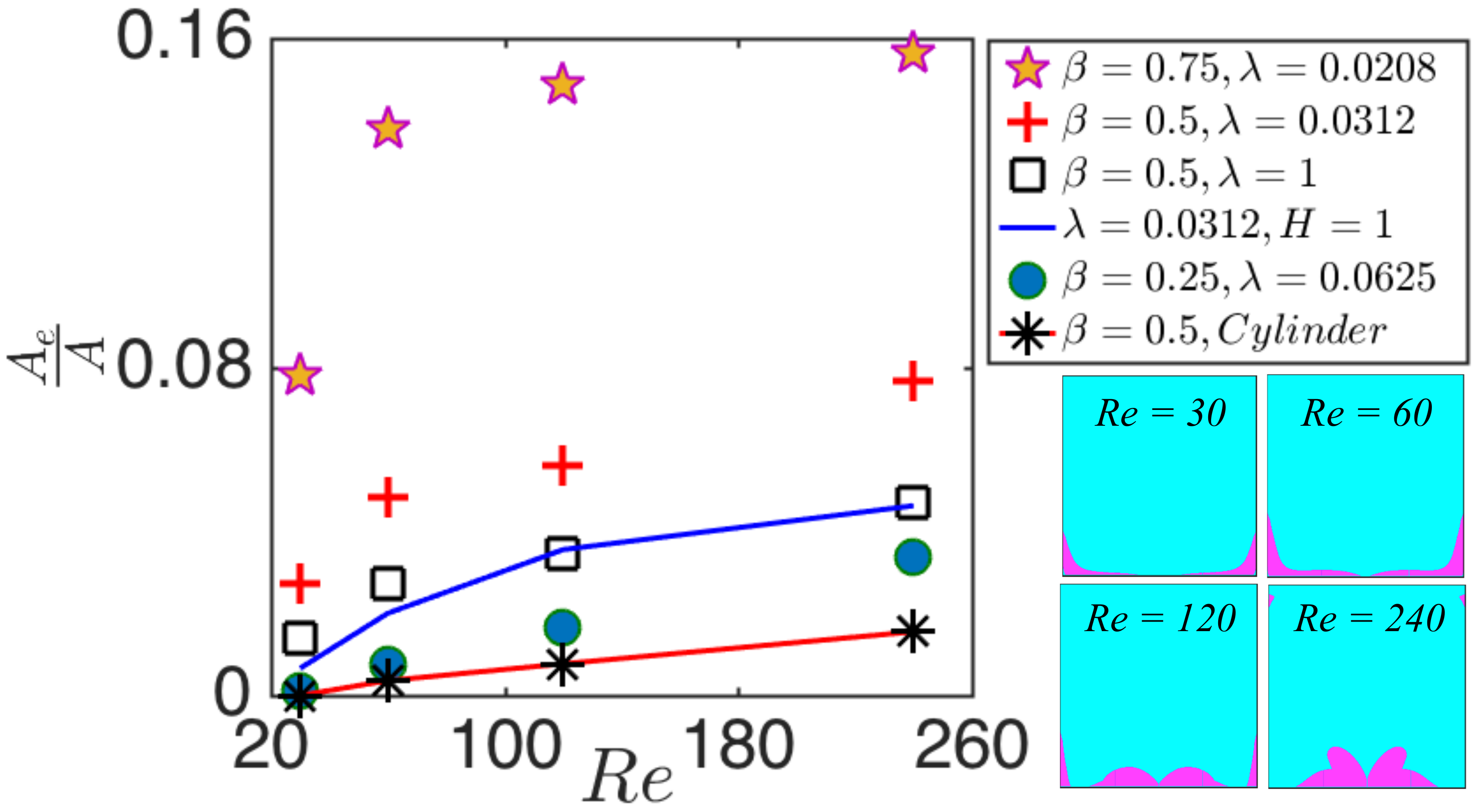}
\caption{The ratio of the entry basin area $A_{e}$ to the total channel cross-sectional area $A$ for all obstacles and channel dimensions. We also show the morphology of the entry basins for the blade obstacle at $Re = 30$, $60$, $120$ and $240$ in this figure. The shape and location of the basins at various $Re$ are the same for all obstacles.   }
\label{fig:Area}
\end{figure}

To discuss the fluid entry more quantitatively, we compute the basin of streamlines that spiral into the vortex behind a $\beta = 0.5$ square cylinder. The size and locations of the entry basins are indicative of the magnitude of fluid exchange with the vortex. In figure~\ref{fig:Basin32}, we exhibit the basin of streamlines in the channel upstream. The streamlines flowing from the entry basins spiral into the vortex. We also show sample streamlines that enter the vortical flow. The basins are computed by following the streamlines starting from a grid of initial points on the $X = 0.03$ plane.  Due to the flow symmetry with respect to $Y = \frac{W}{2}$ (and $Z = \frac{H}{2}$) planes, we show the basins on the upper half of the channel ($Y \geq \frac{W}{2}$).  At $Re = 30$, the basins appear as two narrow ``stripes" adjacent to $Z = 0$ and $H$ walls. With increasing $Re$, additional entry regions form around the channel centerline. At $Re = 240$, the basins appear only around the channel center. The mentioned pattern is observed for all obstacle geometries. \newline

\begin{figure}
\centering
\includegraphics[totalheight=0.28\textheight,]{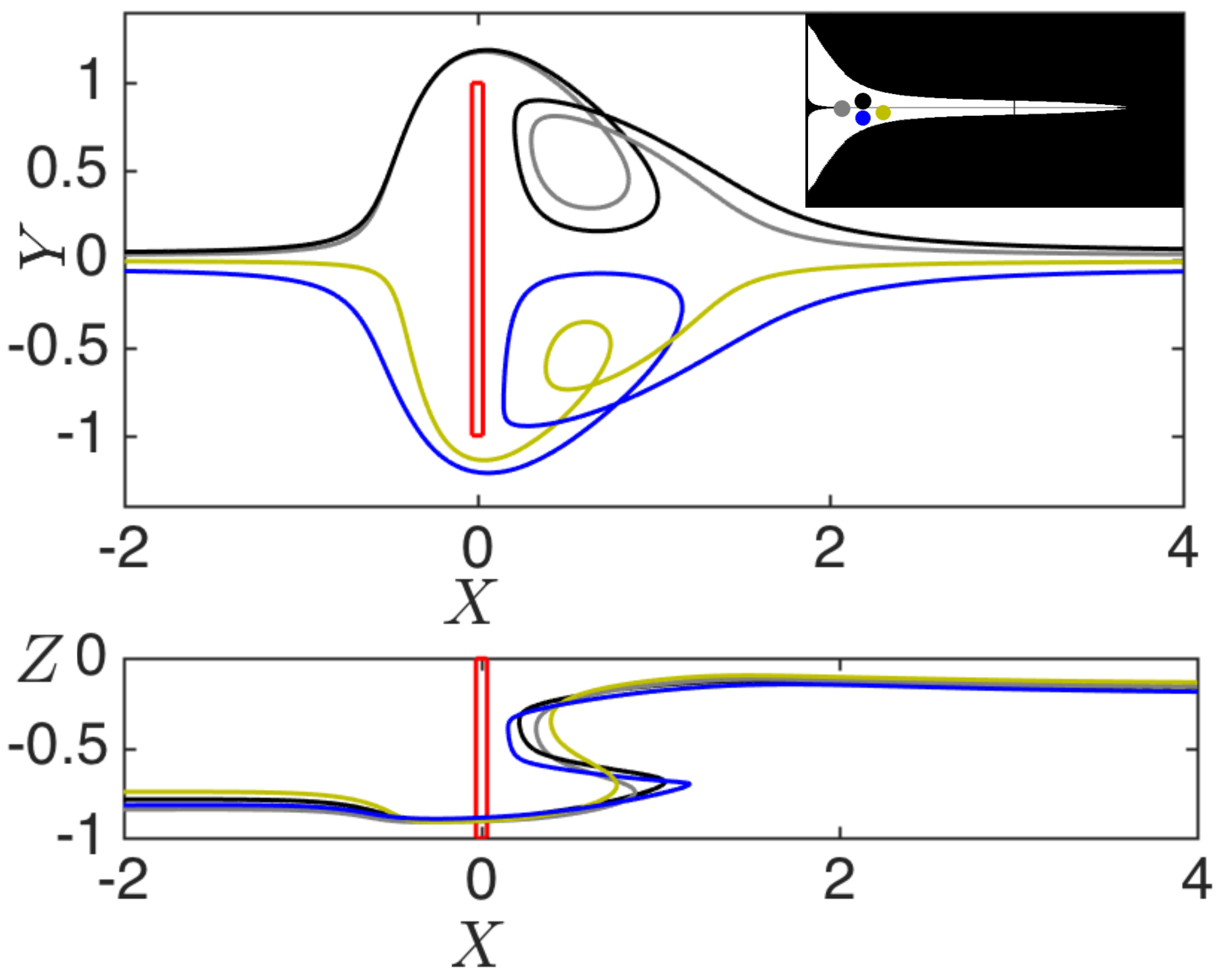}
\caption{Trajectory of isolated particles released in the entry basins at $Re = 30$. The isolated particles enter the vortex behind the $\beta = 0.5, \lambda = 0.0312$ blade cylinder; implying the effect of flow spirals in augmentation of particle entry. }
\label{fig:Sim-Fill}
\end{figure}


The channel height, the obstacle aspect ratio $\lambda$, the blockage ratio $\beta$ and $Re$ affect the size of entry basins. We compare the effect of $\lambda$, the blockage ratio $\beta$ and the channel height by computing the ratio of the basin area $A_{e}$ to the total channel cross-sectional area $A$ for $30 \geq Re \leq 240$. We exhibit the results in figure~\ref{fig:Area}. For all obstacles, the fluid entry is amplified at higher $Re$. The blade obstacle augments the spiraling flow into the vortex. The fluid entry increases further at higher blockage ratios of the blade obstacle (compare $\beta = 0.75 \lambda = 0.0208$, $\beta= 0.5, \lambda = 0.0312$ and $\beta= 0.25, \lambda = 0.0625$). Entry basins do not form in the absence of vortex behind a circular at $Re = 30$, corroborating that the vortex behind the obstacle in a microchannel is formed by the continuous spiraling flow of fluid.  At identical $u_{max}$ (maximum inlet velocity) and the obstacle size, the fluid entry reduces for lower channel heights.   \newline

\begin{figure}
\centering
\subfigure[$Re = 30$]{\includegraphics[totalheight=0.145\textheight,]{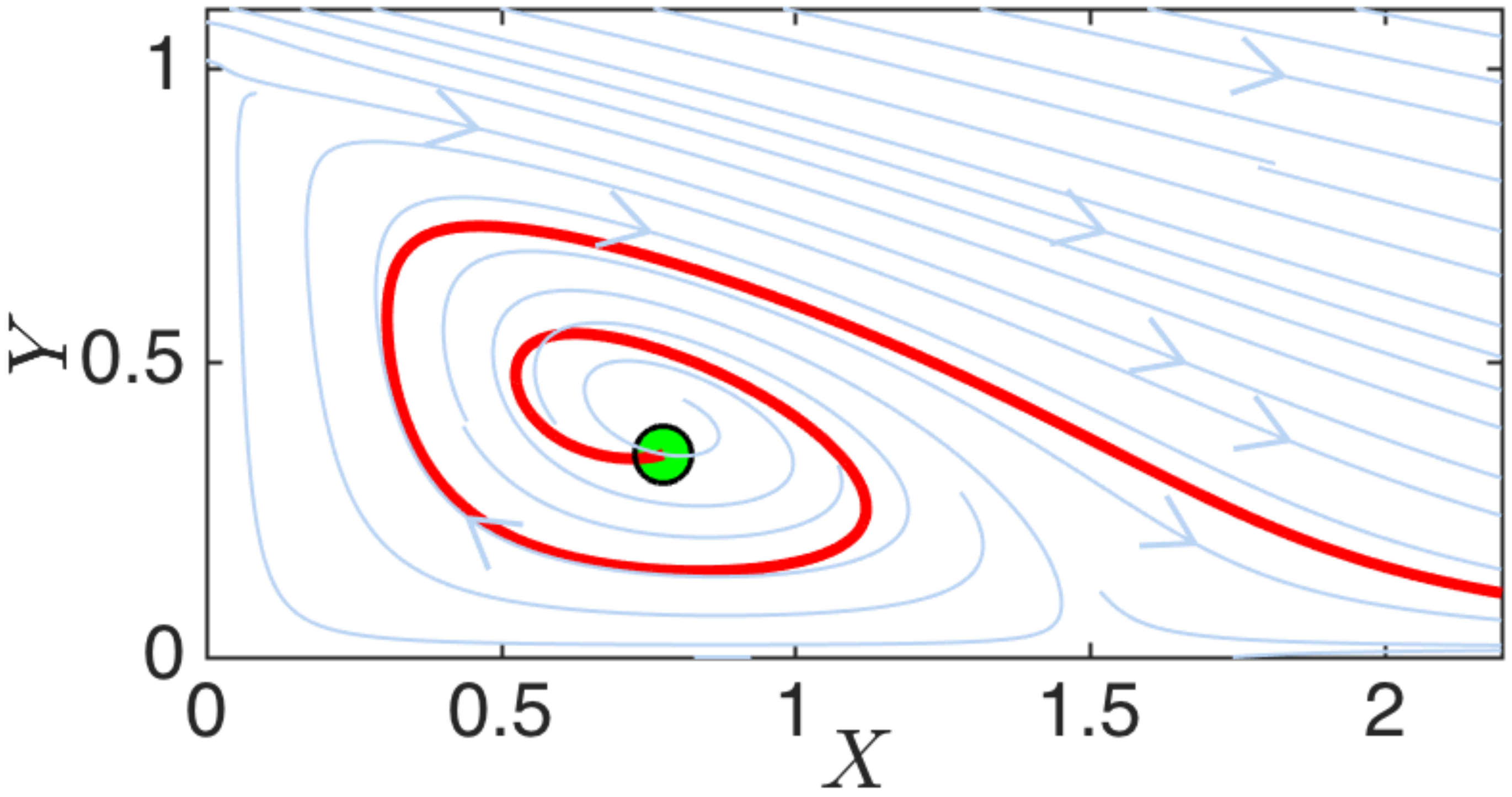}}
\subfigure[$Re = 60$]{\includegraphics[totalheight=0.15\textheight,]{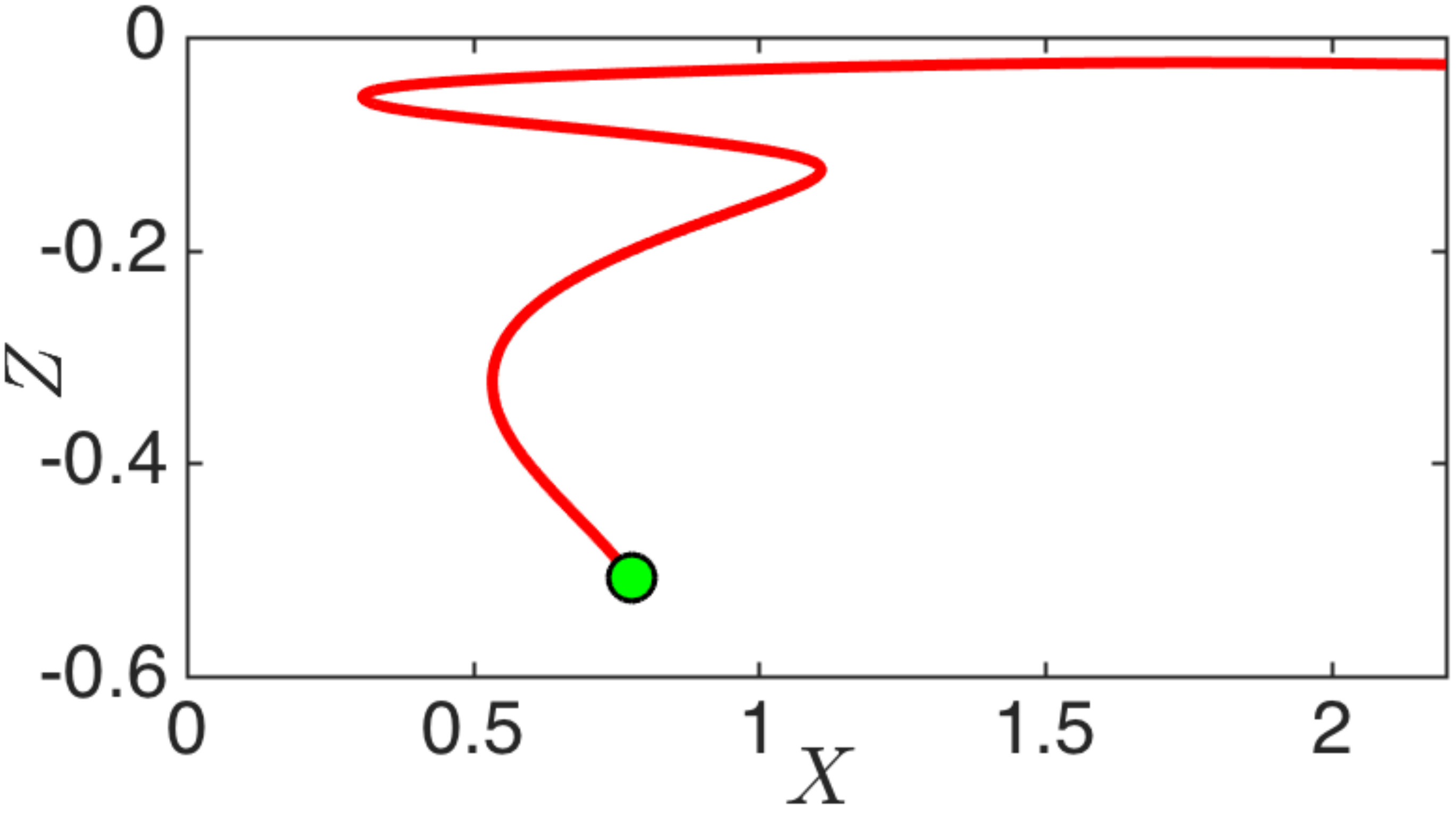}}
\caption{Trajectory of a particle released inside the vortex behind a square at $Re = 60$. The particle trajectory is projected on $XY$ and $XZ$ planes. As opposed to uniform flow where the particle approaches a limit cycle orbit and stays inside the recirculating flow, the particle in the microchannel flow spirals outward on a three-dimensional trajectory towards the middle $Z$ plane and leaves the vortex.  }
\label{fig:Limits}
\end{figure}

\section{Particle exchange with vortices}
In this section, we study flow of a dilute suspension of hard spheres around obstacles in microchannels. Due to small length scales in a microchannel, the particle and channel sizes are comparable and the particle is considered ``finite size". The motion of ``finite size" particles in a microchannel is therefore influenced by hydrodynamic contributions to the particle acceleration.  According to the representation of the Basset-Boussinesque-Oseen (BBO) equation of motion, the additional hydrodynamic contributions on the acceleration of a finite size non-deformable particle are drag, added mass and Basset history forces (Maxey \& Riley 1983). All inertial contributions on the particle force are included in a $\frac{d\bf{V}}{dt}$ term in the BBO equation. Separate effects of inertial contributions such as the inertial lift and added mass on $\frac{d\bf{V}}{dt}$ is not yet clear. The hydrodynamic force of disturbance flow leads to significant deviations from streamlines. In particular, inertial particles migrate towards walls (Ho \& Leal 1974; Di Carlo \emph{et al.} 2009), orbit on stable trajectories inside vortices (Haddadi \& Di Carlo 2017) and bounce back after collision with walls (Vigolo \emph{et al.} 2013). Here we focus on the exchange of particles with the vortex. Considering that particle entry is the collective effect of all hydrodynamic contributions, we do not solely relate the particle entry to the spiraling flow. However, based on topology of the particle trajectory we can implicitly argue the effect of spirals in the background flow on the entry of particles into the vortex.  \newline

\begin{figure}
\centering
\subfigure[$$]{\includegraphics[totalheight=0.171\textheight,]{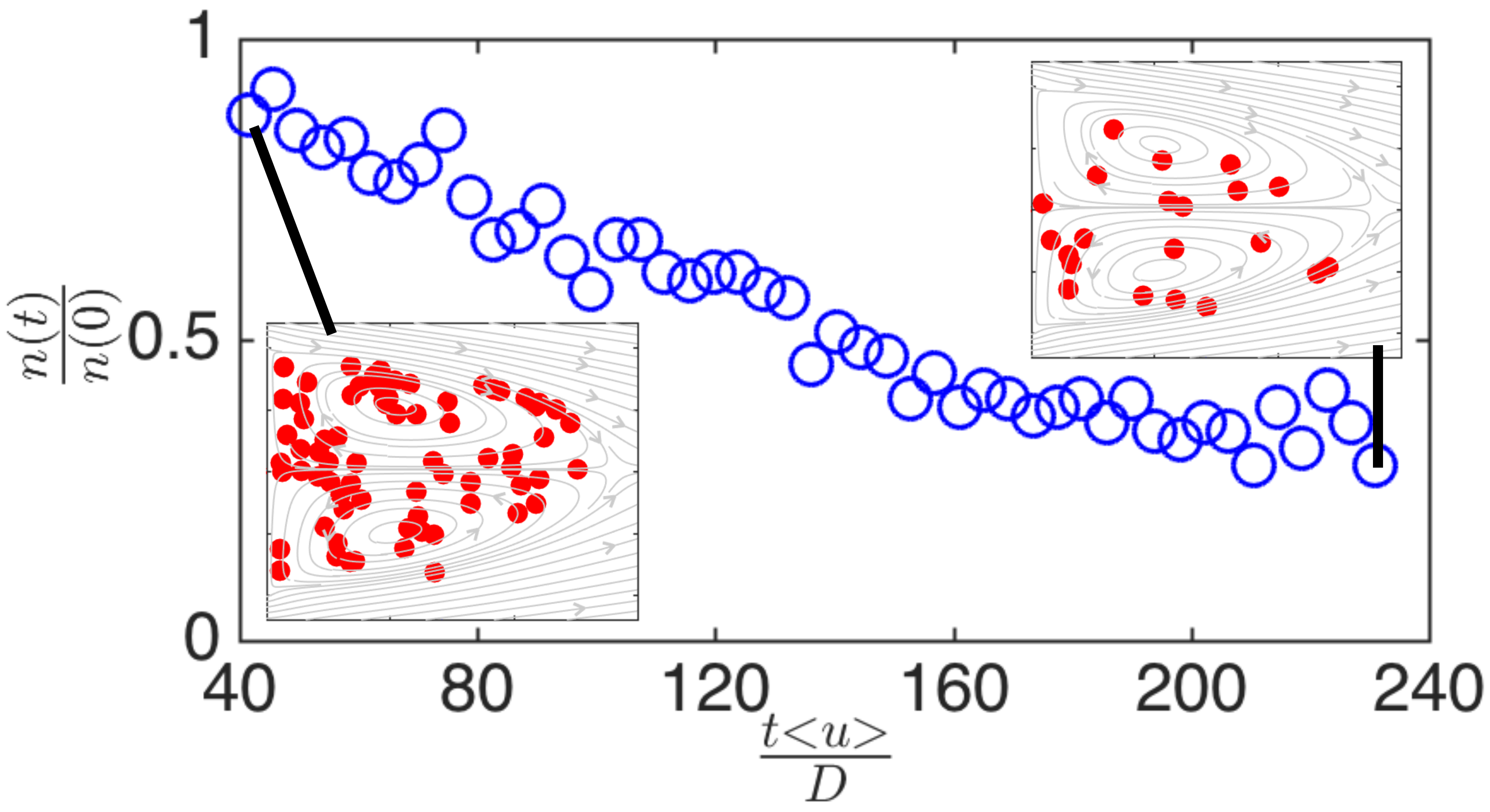}}
\subfigure[$$]{\includegraphics[totalheight=0.171\textheight,]{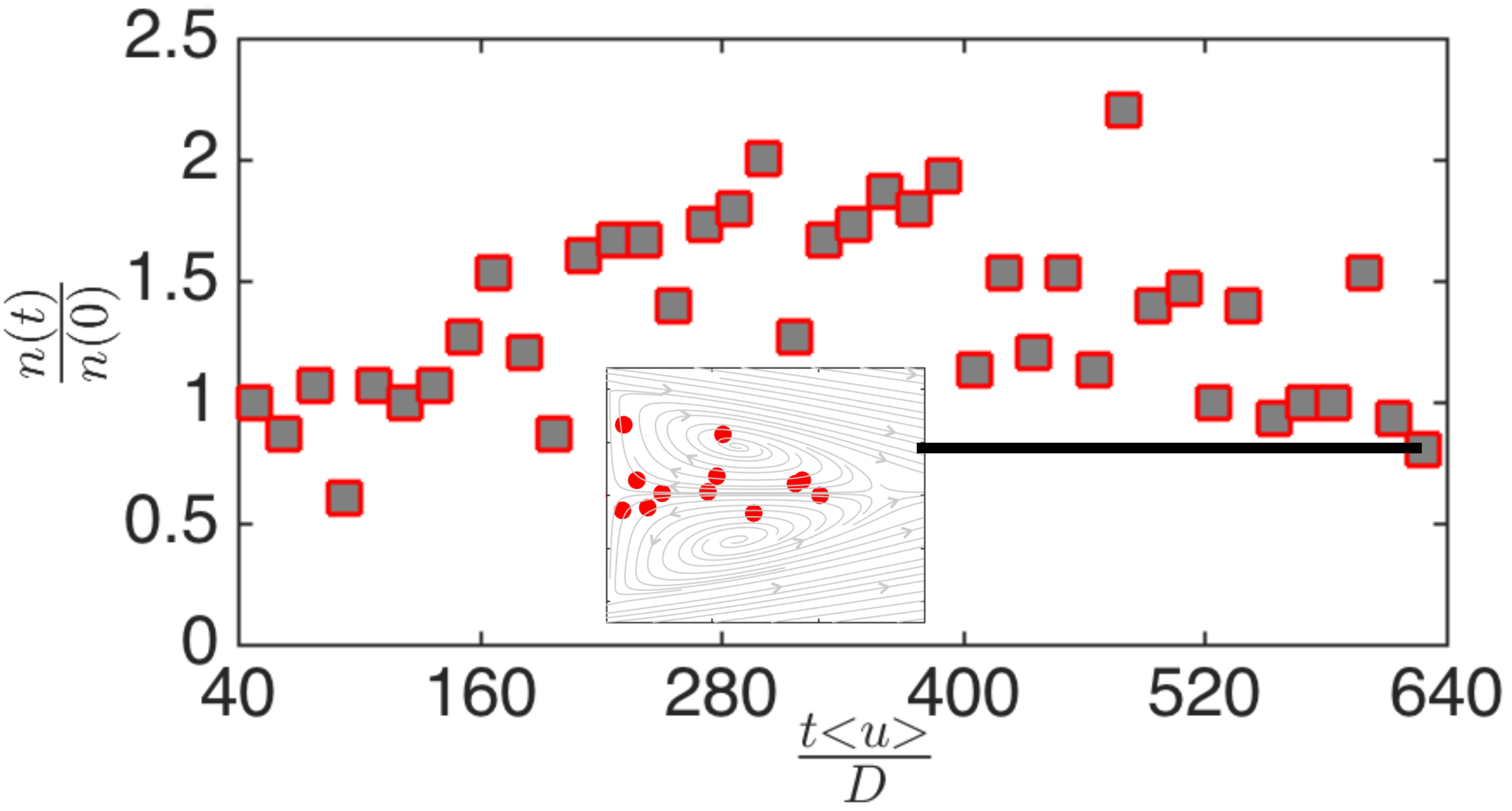}}
\caption{Flow of $\phi = 0.06$ suspension around a square cylinder in (\emph{a}) uniform flow and (\emph{b}) pressure-driven microchannel flow. Considering the difference in fluid velocity profiles, the comparisons are between the vortices with identical size.  }
\label{fig:Boundedness}
\end{figure}

We begin be presenting the trajectory of isolated particles released in the channel upstream. Our previous results of uniform flow around a circular cylinder showed that an isolated particle released in the free stream does not enter the recirculating wake. The particle entry in uniform flow is caused by collision with other particles in a suspension. Figure~\ref{fig:Sim-Fill} shows trajectory of $\delta = \frac{2d}{D_{y}} = 0.1821$ ($d = 5.848$ l.u) particles released in the microchannel upstream in flow around the $\beta = 0.5$ blade at $Re = 30$.  We project trajectories on $XY$ and $XZ$ plane. We also show the initial location of particles on the streamline entry basins. As opposed to uniform flow, an isolated particle released from the basins enters the vortex behind the obstacle. Although the particle acceleration is affected by other hydrodynamic contributions such as inertial lift forces, the similarity between the topology of particle trajectory and the flow spirals (figure~\ref{fig:Entry}) implies the pronounced effect of spiraling flow on the entry of finite size particles into the vortical flow.  \newline


\begin{figure}
\centering
\subfigure[$$]{\includegraphics[totalheight=0.171\textheight,]{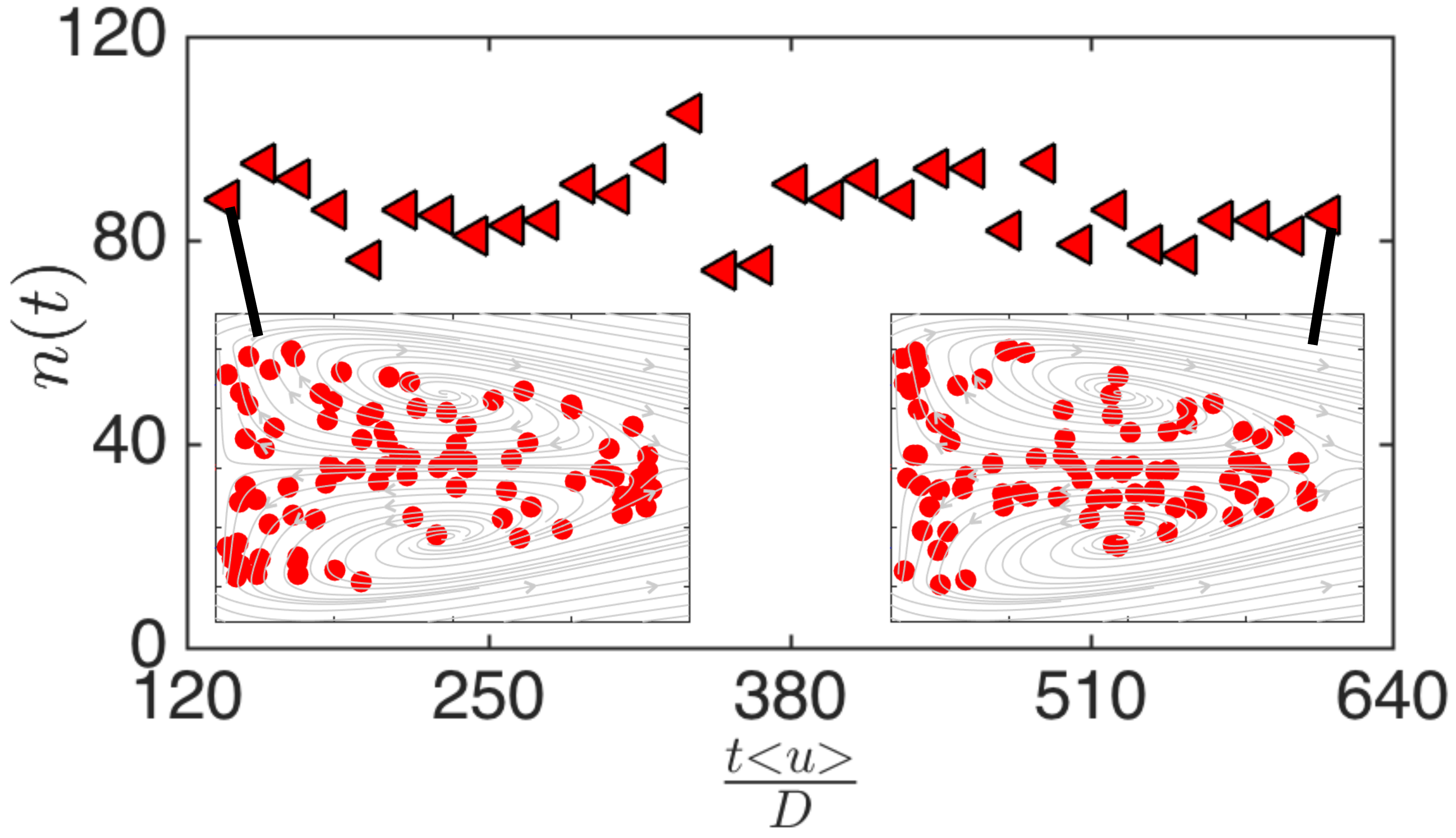}}
\subfigure[$$]{\includegraphics[totalheight=0.171\textheight,]{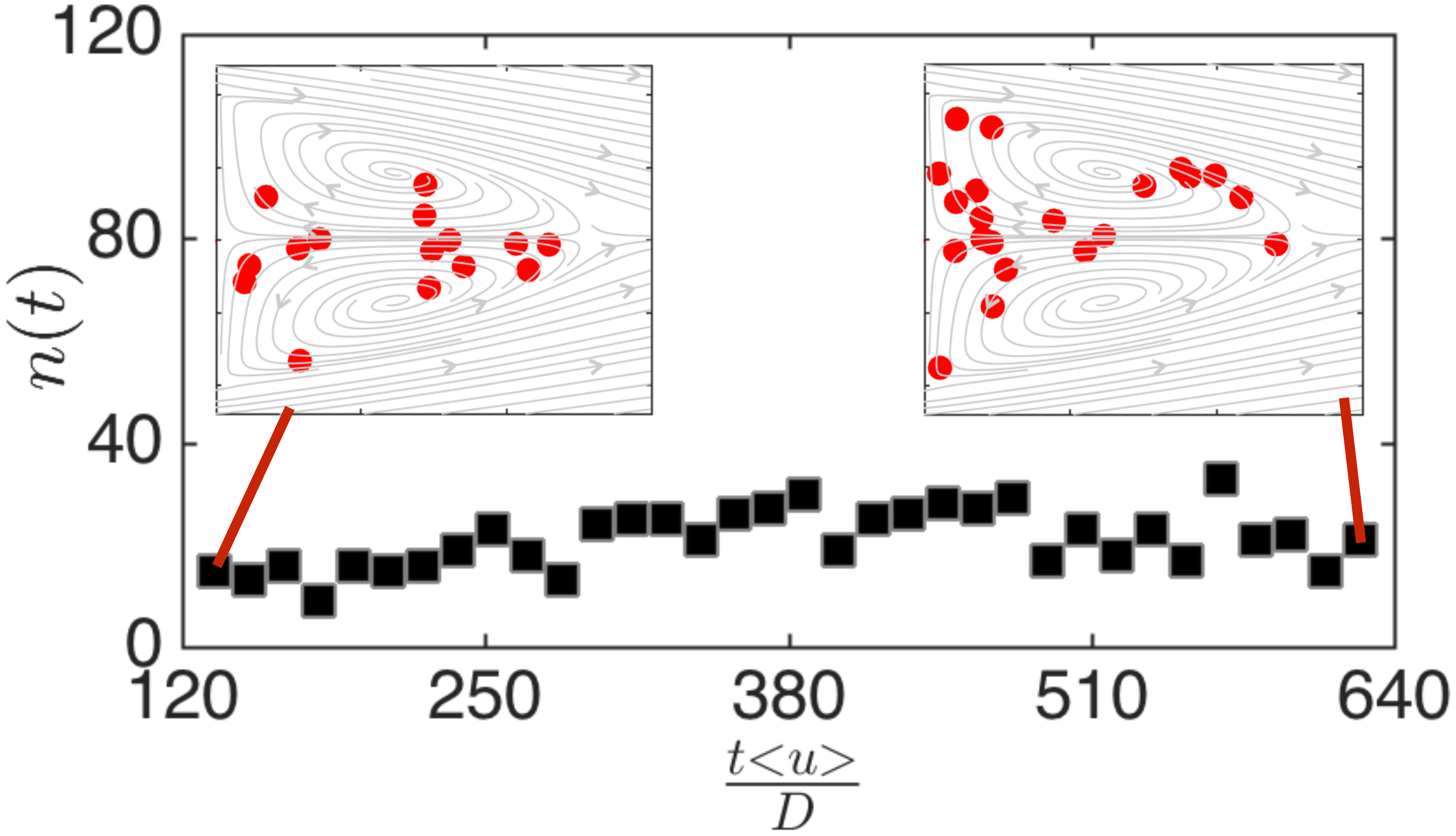}}
\caption{The number of particles in the vortex behind a (\emph{a}) blade and (\emph{b}) square cylinders at $Re = 120$. The vortex behind the blade cylinder entraps significantly larger number of particles. For both obstacles, the invariance of $n(t)$ and time indicates a continuous entry and exit of particles from the vortex.    }
\label{fig:Shape}
\end{figure}

Now we discuss trajectory of particles released inside the vortex. For uniform unbounded flows, we showed that isolated particles trapped in the recirculating flow region migrate towards a stable limit cycle trajectory close to the wake boundaries (Haddadi \emph{et al} 2014; Haddadi, Shojaei-Zadeh \& Morris 2016). An isolated particle orbits on a planar limit cycle without deviation. In figure~\ref{fig:Limits} we exhibit trajectory of an isolated particle released close to the center of the vortex behind a square at $Re = 60$. The particle is released after flow reaches a steady-state. The particle released inside the vortex moves on a three-dimensional spiraling trajectory and exits the wake region. Projections of the particle trajectory on $XY$ and $XZ$ planes show topological similarity of the particle trajectory with the spiraling flow. \newline

To study the continuous exchange of particles with the vortex, here we examine flow of a $\phi = 0.06$ suspension around the obstacle. First we exhibit the time-dependent number of particles inside the wake behind a square obstacle in uniform unbounded flow $~\ref{fig:Boundedness}$(\emph{a}) and microchannel flows (\emph{b}). Considering that the fluid velocity is different in uniform and microchannel flows, we compare the exchange of particles for wakes with identical longitudinal size. We start monitoring the number of particles $n(t)$ by discarding the initial $\frac{t<u>}{D} = 100$ times to remove transient flow effects ($<u>$ is defined differently for uniform and pressure-driven flows). The number of particles in the wake zone $n(t)$ is normalized by $n(t = 0)$, which are $n(0) = 72$ and $15$ for uniform and microchannel flows, respectively. In unbounded flows, $n(t)$ decreases from $n(0) = 72$ to a significantly lower number of $n(t = 231) = 21$ to form a depleted region behind the obstacle. However, $n(t)$ inside the microfluidic vortex does not significantly change. The outward inertial motion of particles inside the wake in both flows leads to formation of wakes with depleted core regions. Similar to suspension flow around the obstacle in uniform unbounded flows, the collision between particles can also increase the entrapment. \newline


We further examine the effect of obstacle shape on flow spirals by comparing $n(t)$ in the vortex behind blade and square cylinders in microchannel flows. We explained that the blade cylinder augments the fluid entry into the vortical flow region. Figure~\ref{fig:Shape} presents $n(t)$ inside the vortex behind the blade (\emph{a}) and square cylinders (\emph{b}). The number of particles inside the vortex behind a blade ($n \approx 80$) is significantly larger than a square cylinder ($n \approx 15$). The vortices do not show a pronounced depletion even after significantly long times of $\frac{t<u>}{D} = 640$. Particles are dispersed uniformly inside the vortex behind the blade. However, simulation snapshots of the vortex behind the square cylinder show formation of a depleted region in the vortex core. Considering that trapped particles move outward and leave the vortex, invariance of $n(t)$ from time is indicative of the continuous entry of particles into the vortex. \newline

\section{Experimental observations}
Before presenting our experimental observations, we state key findings obtained from simulations. In this section we further corroborate these findings by experiments: \newline

(\emph{a}) From studies of fluid exchange with the vortex, we observed that small aspect ratio of the obstacle and $Re$ amplify the entry of fluid into the vortical flow. The blade obstacle enhances the fluid entry compared to an obstacle with square cross-section. \newline

(\emph{a}) Although separate effects of inertial forces on the collective particle acceleration is not yet clear, topological similarity between particle trajectories to flow spirals show the pronounced effect of background flow on the entry of particles. While isolated particles in uniform unbounded flows pass around the obstacle without entry, an isolated particle can be trapped inside the vortex in the microchannel flows. \newline

(\emph{b}) Following the above mentioned remark, we observed the effect of enhanced fluid entry on the presence of more particles in the vortex behind the blade cylinder. For both obstacles, the number of particles inside the vortex does not change, showing a continuous entry and exit of particles into the vortex. Less particle entry combined with outward spiraling motion result in formation of an empty core region in the vortex behind the square cylinder.   \newline

In order to further demonstrate the above mentioned remarks, here we present results of our microchannel experiments. The effect of $\lambda$ and $Re$ on the particle entry can be observed from the state of vortex behind the $\beta = 0.5$ blade and square cylinders. The flow is visualized by fluorescent imaging of $d = 1\mu$m tracers. The microchannels are fabricated using a standard soft-lithography procedure used in our previous work (Haddadi \& Di Carlo 2017). Due to inherent errors in the soft-lithography procedure, we chose higher aspect ratio of $\lambda = 0.12$ to fabricate a blade obstacle in our experiments. The channel length $L$, width $W$ and height $H$ are $2$cm, $216 \mu$m and $23 \mu$m respectively. We also used $H = 60 \mu$m channels and obtained same results. Similar to LBM simulations, the effect of reducing $\lambda$ and increasing $Re$ in amplifying the particle entry can be observed in streaklines of tracers. Here we define $Re$ using the hydraulic diameter of the microchannel and the average inlet velocity $<u> $ as $Re_{h} = \frac{<u>h_d}{\nu}$. Both $h_{d}$ and $<u>$ can be accurately measured for a conduit of known size.   \newline

\begin{figure}
\centering
\subfigure[$Re_h = 2.1$ ]{\includegraphics[totalheight=0.09\textheight,]{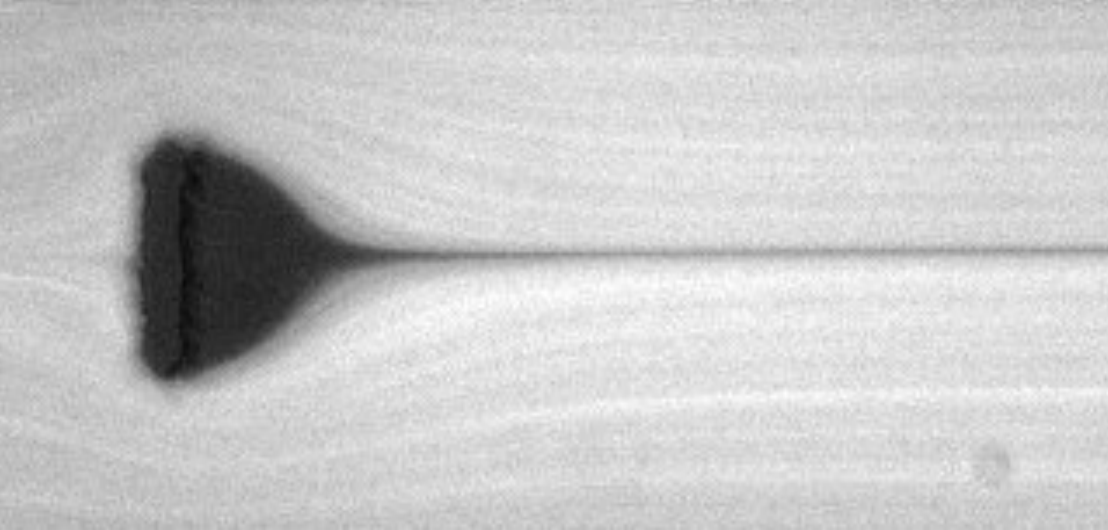}}
\subfigure[$Re_h = 4.4$ ]{\includegraphics[totalheight=0.09\textheight,]{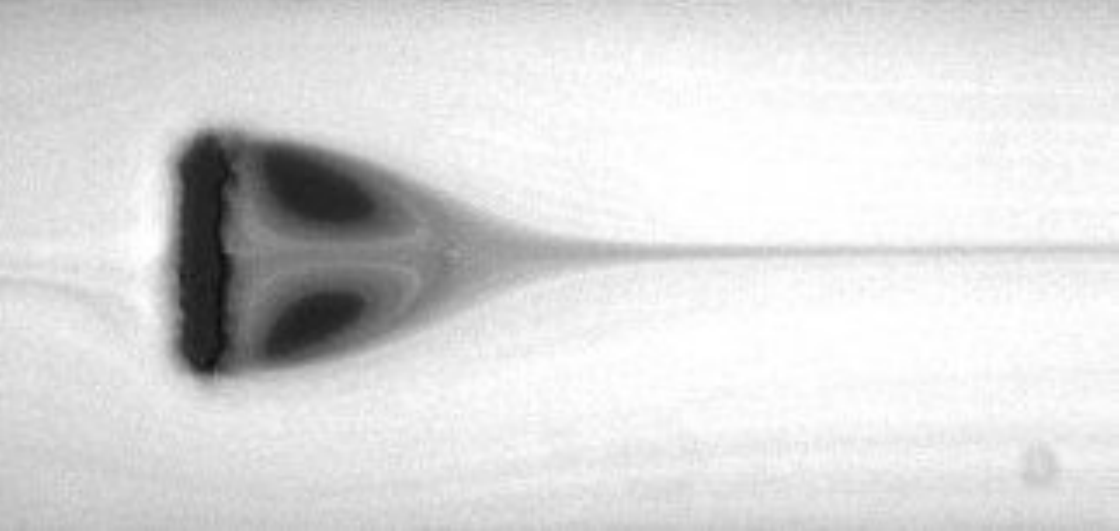}}
\subfigure[$Re_h = 8.8$ ]{\includegraphics[totalheight=0.09\textheight,]{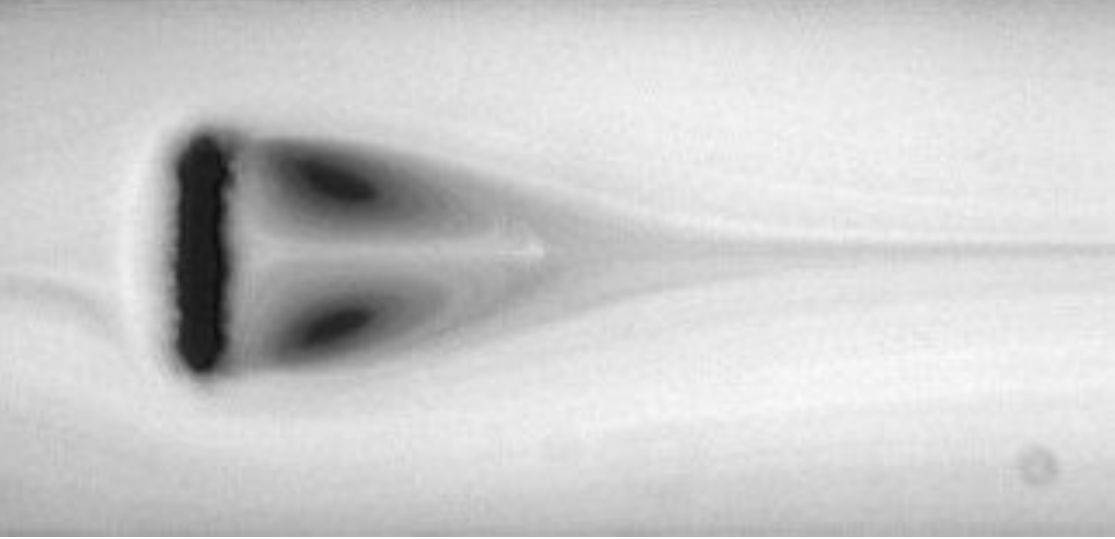}}\\
\subfigure[$Re_h = 2.1$ ]{\includegraphics[totalheight=0.09\textheight,]{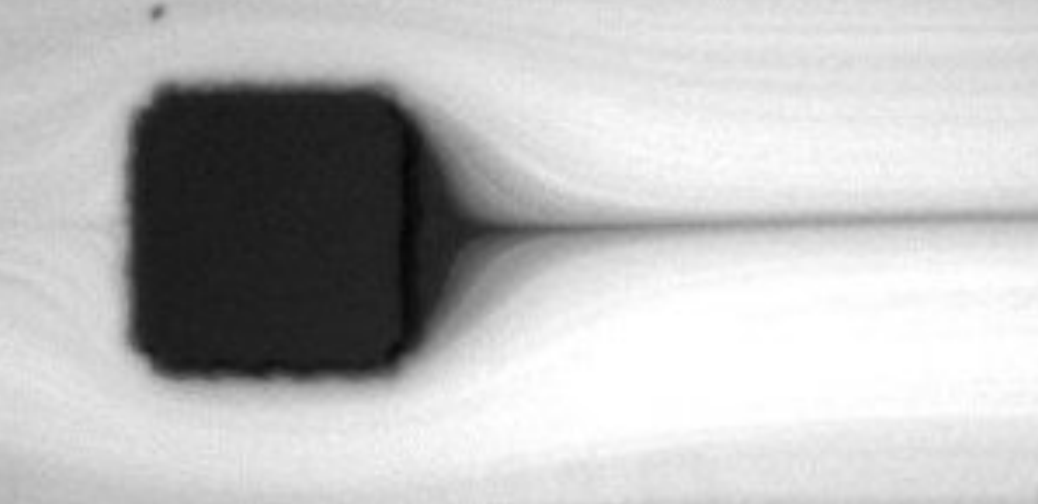}}
\subfigure[$Re_h = 4.4$ ]{\includegraphics[totalheight=0.09\textheight,]{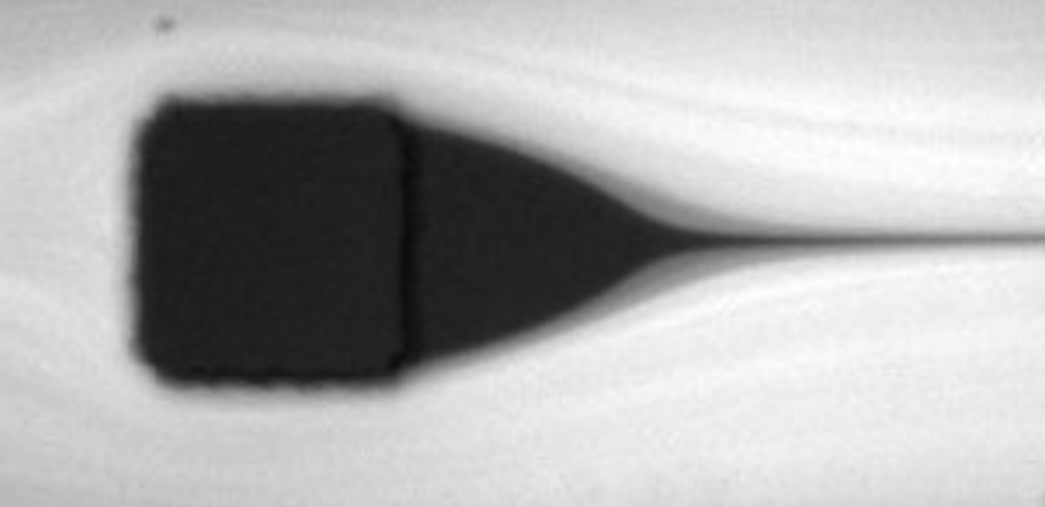}}
\subfigure[$Re_h = 8.8$ ]{\includegraphics[totalheight=0.09\textheight,]{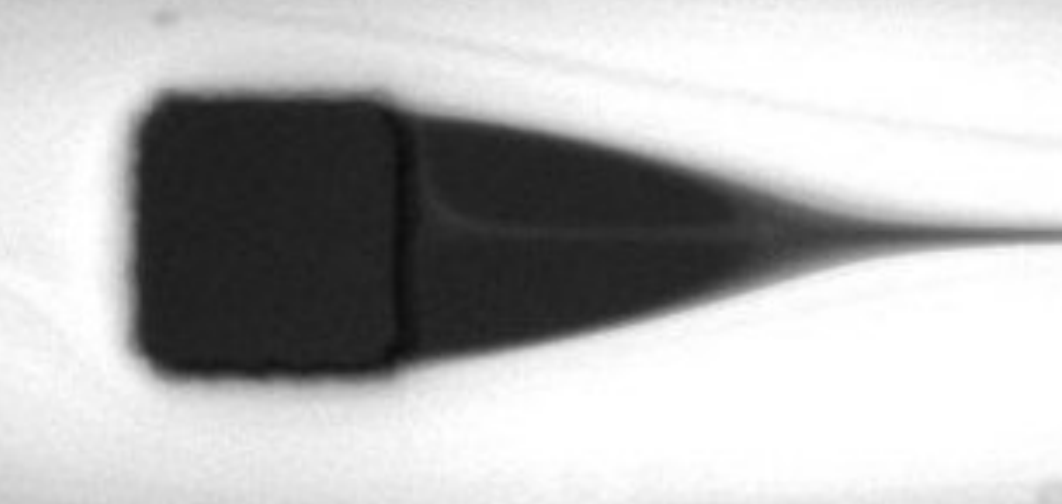}}\\
\caption{Experimental observation of flow around the obstacle in microchannels. The flow is visualized by fluorescent imaging of $d = 1\mu$m tracers. Consistent with numerical results, increasing $Re_{h}$ and reducing $\lambda$ amplifies the fluid entry.  }
\label{fig:O-Shape}
\end{figure}

\section{Concluding remarks}
In this work, we utilized lattice-Boltzmann simulations and microchannel experiments to study the flow of Newtonian fluids and dilute suspensions over obstacles in microchannels. In particular, we highlighted the effect of flow spirals in microchanel flows on the exchange of fluid and particles with a vortical flow behind the obstcale. We observed that increasing the blockage ratio, reducing the obstacle ratio for bluff bodies with rectangular cross-sectional area and $Re$ amplify the entry of fluid into the vortex. The fluid entry leads to continuous exchange of particles with the vortex. For instance, the vortex behind the blade contains significantly larger number of particles compared to a square cylinder. Both fluid and particle exchange are characteristics of inertial flow around obstacles in microchannels. \newline 

\section{Acknowledgement}
I would like to thank my PhD thesis mentor, professor Jeffrey Morris in the City College of New York and my postdoctoral research adviser, professor Dino Di Carlo in University of California Los Angeles for their guidance, instructions and generous support.


\bibliographystyle{jfm-references}

\end{document}